\documentclass[aps,prb,twocolumn
,groupedaddress
,superscriptaddress
,amsfonts,
citeautoscript,
a4paper]{revtex4-2}

\usepackage{upgreek}
\usepackage{mathrsfs}
\usepackage[utf8]{inputenc}
\usepackage[T1]{fontenc}
\usepackage{amssymb,amsmath,bm}
\usepackage[pdftex]{graphicx}
\usepackage{color,soul}
\usepackage{mathtools}
\usepackage{amsmath}
\usepackage[margin=2.6cm]{geometry}
\usepackage{natbib}
\usepackage{xcolor}
\usepackage{hyperref}
\usepackage{listings}
\usepackage{bbold}
\hypersetup{colorlinks=true,linkcolor=blue,citecolor=blue}
\usepackage{orcidlink}

\hypersetup{colorlinks=true,linkcolor=blue,citecolor=blue}

\usepackage{placeins}

\providecommand{\pnl}[1]{{\textcolor{blue}{#1}}}

\let\Re\relax % remove the default definition before redefining
\DeclareMathOperator{\Re}{Re}
\let\Im\relax % remove the default definition before redefining
\DeclareMathOperator{\Im}{Im}

\begin{document}
\title{Nonlocal electrodynamics of two-dimensional anisotropic magneto-plasmons}

\author{A. J. Chaves\,\orcidlink{0000-0003-1381-8568}}
\email{andrejck@ita.br}
\affiliation{Department of Physics, Aeronautics Institute of Technology, 12228-900, São José dos Campos, SP, Brazil}
\affiliation{POLIMA---Center for Polariton-driven Light--Matter Interactions, University of Southern Denmark, Campusvej 55, DK-5230 Odense M, Denmark}

\author{Line Jelver\,\orcidlink{0000-0001-5503-5604}}
\affiliation{POLIMA---Center for Polariton-driven Light--Matter Interactions, University of Southern Denmark, Campusvej 55, DK-5230 Odense M, Denmark}

\author{D. R. da Costa\,\orcidlink{0000-0002-1335-9552}}
\email{diego\_rabelo@fisica.ufc.br}
\affiliation{Departamento de F\'isica, Universidade Federal do Cear\'a, Campus do Pici, 60455-900 Fortaleza, Cear\'a, Brazil}
\affiliation{Department of Physics, University of Antwerp, Groenenborgerlaan 171, B-2020 Antwerp, Belgium}

\author{Joel~D.~Cox\,\orcidlink{0000-0002-5954-6038}}
\affiliation{POLIMA---Center for Polariton-driven Light--Matter Interactions, University of Southern Denmark, Campusvej 55, DK-5230 Odense M, Denmark}
\affiliation{Danish Institute for Advanced Study, University of Southern Denmark, Campusvej 55, DK-5230 Odense M, Denmark}

\author{N.~Asger~Mortensen\,\orcidlink{0000-0001-7936-6264}}
\email{asger@mailaps.org}
\affiliation{POLIMA---Center for Polariton-driven Light--Matter Interactions, University of Southern Denmark, Campusvej 55, DK-5230 Odense M, Denmark}
\affiliation{Danish Institute for Advanced Study, University of Southern Denmark, Campusvej 55, DK-5230 Odense M, Denmark}

\author{Nuno~M.~R.~Peres\,\orcidlink{0000-0002-7928-8005}}
\email{peres@fisica.uminho.pt}
\affiliation{POLIMA---Center for Polariton-driven Light--Matter Interactions, University of Southern Denmark, Campusvej 55, DK-5230 Odense M, Denmark}
\affiliation{Centro de F\'{\i}sica (CF-UM-UP) and Departamento de F\'{\i}sica, Universidade do Minho, P-4710-057 Braga, Portugal}
\affiliation{International Iberian Nanotechnology Laboratory (INL), Av Mestre Jos\'e Veiga, 4715-330 Braga, Portugal}

\date{\today}

\begin{abstract}
We present a hydrodynamic model, grounded in Madelung's formalism, to describe collective electronic motion in anisotropic materials. This model incorporates nonlocal contributions from the Thomas--Fermi quantum pressure and quantum effects arising from the Bohm potential. We derive analytical expressions for the magnetoplasmon dispersion and nonlocal optical conductivity.
To demonstrate the applicability of the model, we examine electrons in the conduction band of monolayer phosphorene, an exemplary anisotropic two-dimensional electron gas. The dispersion of plasmons derived from our hydrodynamic approach is closely aligned with that predicted by \emph{ab~initio} calculations. 
Then, we use our model to analyze few-layer black phosphorus, whose measured infrared optical response is hyperbolic. Our results reveal that the incorporation of nonlocal and quantum effects in the optical conductivity prevents black phosphorus from supporting hyperbolic surface plasmon polaritons. We further demonstrate that the predicted wavefront generated by an electric dipole exhibits a significant difference between the local and nonlocal descriptions for the optical conductivity.
This study underscores the necessity of moving beyond local approximations when investigating anisotropic systems capable of hosting strongly confined plasmon-polaritons.
\end{abstract}

\maketitle

\section{Introduction}

The optical response of crystals is generally anisotropic due to the breaking of rotational symmetry by the discrete crystal lattice. This optical anisotropy gives rise to the phenomenon of birefringence and is crucial for various optical device concepts, owing to their ability to modulate and polarize light~\cite{Weber2000}. Although anisotropic optics is typically discussed in the context of dielectrics~\cite{born2013principles}, metallic systems can also exhibit optical anisotropy in both linear~\cite{furtak1975anisotropic,Tadjeddine1980} and nonlinear responses~\cite{Friedrich:1989,Lupke:1990,Boroviks:2021}. At the core of light-matter interactions in metals or doped semiconductors are plasmons, which can inherit the anisotropic properties of their parent materials~\cite{Tadjeddine1980}.

Recently, the advent of two-dimensional (2D) materials~\cite{Pulickel2016} has introduced inherent anisotropy due to the stark difference between in-plane and out-of-plane responses. For instance, insulating hexagonal-phase boron nitride (hBN) possesses two Reststrahlen bands and hyperbolic phonon polaritons in the infrared~\cite{Li2015}, even at the monolayer limit~\cite{Dai2019}. Additionally, in-plane hyperbolic phonon-polaritons are observed in the semiconductor orthorhombic alpha-phase molybdenum trioxide (\mbox{$\alpha$-MoO$_3$})~\cite{Zheng2018}, which exhibits extremely confined modes with low damping~\cite{Taboada2024}. Monoclinic crystals, such as beta-phase gallium oxide ($\beta$-Ga$_2$O$_3$), can support shear polaritons due to non-orthogonal principal crystal axes~\cite{Passler2022}. The phonon modes of low-symmetry crystals can hybridize with plasmons, as seen in hybridized surface plasmon-phonon polaritons in hBN/graphene heterostructures~\cite{kumar2015tunable} and MoO$_3$ over a gold (Au) substrate~\cite{Barcelos2021}. However, only the latter exhibits anisotropic plasmon-phonon dispersion due to the in-plane anisotropy of MoO$_3$. Low-symmetry van~der~Waals doped semiconductors, such as black phosphorus (BP), and 2D metals like molybdenum chloride dioxide (MoOCl$_2$), can exhibit innate anisotropic plasmons~\cite{Liu2016,Venturi2024,Ruta:2025}.

One of the primary applications of surface polaritons is their ability to confine light in small volumes, or equivalently, possess a high wavenumber compared to free-space radiation of the same frequency~\cite{maradudin2014modern}. In-plane hyperbolic polaritons can exceed the limits of elliptic ones due to their unbounded isofrequency dispersion relation in reciprocal space, shaped as a hyperbola~\cite{Wang2024}. However, in the regime of high wavenumbers, nonlocal electrodynamics cannot be disregarded~\cite{Goncalves2020,Mortensen:2021,monticone2025}. Thus, nonlocal effects are expected to be significant in describing the optical properties of surface polaritons. This conclusion has been supported by experimental observations of propagating gap surface plasmon modes in ultrathin metal–dielectric–metal planar waveguides~\cite{Boroviks2022}, where the high wavenumber of graphene plasmons was used to access the nonlocal response of metals. Similarly, nonlocal effects have been observed in 2D plasmons in graphene-on-metal systems~\cite{Lundeberg:2017,Dias:2018}.

The nonlocal response of materials can be determined using the random-phase approximation (RPA), such as the Lindhard response function or the Kubo formula for the optical conductivity, considering transitions with finite energy and momentum. These calculations can be performed within an \emph{ab~initio} framework~\cite{Correas-Serrano_2016}. Another similar approach is time-dependent density functional theory (TDDFT)~\cite{Ullrich2011}. However, for mesoscopic problems, such as nanostructures containing thousands of atoms, these approaches become unfeasible, necessitating further approximations. One such model is the quantum hydrodynamic model (QHM), which can capture both qualitative and quantitative aspects of nonlocality. It associates the electron liquid with a local density $n(\mathbf{r})$ and a velocity field $\mathbf{u}(\mathbf{r})$, governed by the compressible Euler equations where the internal pressure has a quantum origin. One approximation for the quantum pressure is the Thomas--Fermi QHM, where the pressure is given by the degeneracy pressure of a free Fermi gas. More sophisticated approximations have been developed using density functional theory (DFT) exchange and correlation functionals~\cite{ciraci2017}. The hydrodynamic model has been extensively used since the 1970s to discuss surface effects in metals~\cite{Barton_1979} and more recently to describe plasmons in noble metals~\cite{Chen2023}, alkali metals~\cite{Teperik2013}, heavily doped semiconductors~\cite{Ciraci2021}, and 2D materials such as graphene~\cite{Chaves2017} and twisted bilayer graphene~\cite{Barbosa2025}. The model has also been used to study nanostructures, including Au nanoparticles~\cite{Ciraci2012} and plasmonic gap structures~\cite{Zhang2025}.

In the case of graphene, an isotropic 2D material, the hydrodynamic model has been used to describe the propagation of plasmonic wakes due to the drag of charged particles moving parallel to a graphene sheet~\cite{Chaves2017}, in the study of terahertz laser combs~\cite{Cosme2020}, and double-layer structures~\cite{Hua2025}. In the presence of a perfect conductor, the nonlocal correction in the QHD becomes increasingly important as the distance between the graphene layer and the conductor decreases~\cite{Moradi2023}. However, there is limited literature on anisotropic plasmons~\cite{Seongjin2021}. Building on our recent Madelung considerations for in-plane isotropic 2D systems~\cite{Cardoso:2025}, this manuscript aims to address the effects of nonlocality on anisotropic plasmons using hydrodynamic equations.

The paper is organized as follows: in Sec.~\ref{sec:Model} we derive the hydrodynamic equations for anisotropic systems using the Madelung formalism, obtaining the dispersion relation for anisotropic plasmons and discussing the magnitude order for the plasmon wavelength that impacts each. In Sec.~\ref{sec:applications} we apply the formalism to characterize the optical response of doped BP, an anisotropic semiconductor. We start this section by describing first-principle calculations \ref{subsec:abinitio} that were used to compare with Madelung's hydrodynamic approach~\ref{subsec:comparison}. Then, we discuss the magnetoplasmon dispersion relation and how nonlocal effects impact the velocity field~\ref{subsec:magnetoplasmons}. In Subsec.~\ref{subsec:spp} we discuss the effects of nonlocality on the surface plasmon-polariton spectrum, revealing that nonlocal effects inhibit the appearance of hyperbolic modes. In Subsec.~\ref{subsec:purcell}, plasmonic wakes due to oscillating electric dipoles and the Purcell effect are addressed. Finally, in Sec.~\ref{sec:ending}, our conclusions and perspectives are presented.

\section{Hydrodynamic model for anisotropic materials} \label{sec:Model}

\subsection{Madelung-like derivation of the anisotropic hydrodynamic model}

Building on Madelung’s seminal work~\cite{madelung1927quantum}, which reformulated the Schrödinger equation as a hydrodynamic model, we adopt a similar approach to describe the electromagnetic response of an anisotropic two-dimensional electron gas (2DEG). In this framework, we derive a hydrodynamic formulation based on the continuity and Euler equations. As our starting point, the corresponding anisotropic time-dependent Schr{\"o}dinger equation reads
\begin{equation}\label{eq:schrodinger}
i\hbar \partial_t \Psi=-\frac{\hbar^2}{2} \sum_{j=x,y} \frac{1}{m_j}
\partial_j^2 \Psi+U\Psi,
\end{equation}
where $m_x$ and $m_y$ are the effective masses along the $x$ and $y$-directions, respectively, and $U$ is the potential energy. Stationary states are associated with a wavefunction of the form $\Psi(\mathbf{r},t)=\psi(\mathbf{r}) e^{-iEt/\hbar}$. Madelung's approach consists in assuming that a general wavefunction $\Psi$, not necessarily a stationary state, can be written as
\begin{equation}\label{eq:psi_madelung}
\Psi(\mathbf{r},t)=\alpha(\mathbf{r},t)e^{i\beta(\mathbf{r},t)}, 
\end{equation}
where $\alpha$ and $\beta$ are real functions depending both on the spatial and temporal coordinates $\mathbf{r}$ and $t$. Substituting Madelung's trial wavefunction~\eqref{eq:psi_madelung} into Schr{\"o}dinger's equation~\eqref{eq:schrodinger}, one obtains for the real and imaginary equation's parts, respectively, that
\begin{subequations}
\begin{align}
&\alpha \hbar\partial_t\beta=\sum_{j=x,y} \hspace{-0.05cm}\frac{\hbar^2}{2m_j}
 \left[\partial_j^2\alpha -\alpha\left(\partial_j\beta\right)^2\right] - \alpha U, \label{eq:real}\\
& \hbar \partial_t \alpha = - \sum_{j=x,y} \hspace{-0.05cm} \frac{\hbar^2}{2m_j} \left(\alpha\partial_j^2\beta+2\partial_j\alpha\partial_j\beta\right) , \label{eq:imag}
\end{align}
\end{subequations}
where the dependence of $\alpha$ and $\beta$ on $\mathbf{r}$ and $t$ is implicit. By defining the field $\mathbf{v} = \mathbf{\nabla}_\parallel\beta$, and multiplying Eq.~\eqref{eq:imag} by $\alpha$, we can transform the imaginary part of the equation into 
\begin{equation}
\partial_t \alpha^2 + \hbar\sum_{j=x,y} \frac{1}{m_j} \left(\alpha^2 \partial_jv_j+ v_j\partial_j\alpha^2\right)=0,
\end{equation}
which can in turn be written as a continuity equation
\begin{equation}\label{eq:continuity}
\partial_t n+ \mathbf{\nabla}_\parallel\cdot\left(n \mathbf{u}\right)=0,
\end{equation}
with $u_j= \hbar v_j/m_j$ having units of velocity and $n=\alpha^2$ being the electronic density.

By analyzing the real part of the Schr{\"o}dinger equation, one can derive Euler's equation. To do so, we divide Eq.~\eqref{eq:real} by $\alpha$ and apply the gradient-like operator $m_k^{-1}\partial_k$, resulting in
\begin{equation}\label{eq:real3}
\partial_t u_k \hspace{-0.05cm}+\hspace{-0.05cm} \frac{\partial_k}{ m_k} \hspace{-0.12cm} \sum_{j=x,y} \hspace{-0.12cm} \frac{m_j u_j^2}{2} \hspace{-0.05cm} = \hspace{-0.05cm}\frac{\hbar^2}{2} \hspace{-0.12cm}\sum_{j=x,y} \hspace{-0.12cm}\frac{\partial_k}{m_j m_k} \hspace{-0.1cm}\left(\hspace{-0.05cm}\frac{1}{\alpha}\partial_j^2\alpha \hspace{-0.05cm}\right) \hspace{-0.05cm}-\hspace{-0.05cm} \frac{\partial_k U}{m_k}. 
\end{equation}
Here, the term proportional to $\hbar^{2}$ on the right-hand side (RHS) is the Bohm potential, the second term of the RHS has units of force per mass, and the second term of the left-hand side (LHS) can be viewed as the spatial derivative of the kinetic energy. By dimensional analysis of the terms in Eqs.~\eqref{eq:continuity} and \eqref{eq:real3}, one notices that $\alpha^{2}$ can be interpreted as a density $n$ and $\alpha^{2}\mathbf{u}$ as a current density of particles.

The anisotropic Bohm potential,
\begin{equation}
\label{eq:Bohm-potential}
    V^B=\frac{\hbar^2}{2} \hspace{-0.12cm}\sum_{j=x,y} \hspace{-0.12cm}\frac{1}{m_j } \hspace{-0.1cm}\left(\hspace{-0.05cm}\frac{1}{\sqrt{n}}\partial_j^2\sqrt{n} \hspace{-0.05cm}\right) \hspace{-0.05cm},
\end{equation}
originates from the kinetic term of the Schr{\"o}dinger equation. Unlike a typical local potential, which depends only on $\mathbf{r}$, the spatial derivative here introduces a source of nonlocality, extending its influence to the near vicinity of $\mathbf{r}$.

\subsection{Anisotropic 2D Euler equation}

We will include the electromagnetic field as a classical field in the electrostatic approximation, i.e., the electric potential $\Phi$ is the solution of the Poisson equation,
\begin{equation}
\mathbf{\nabla}^2\Phi= \frac{1}{\epsilon_0} \left[ \rho_\mathrm{ext}- e(n-n_0)\delta(z) \right], \label{eq:Poisson}
\end{equation}
with $\varepsilon_0$ being the vacuum permittivity, $z$ being the axis transverse to the 2D material sheet, and $\rho_\mathrm{ext}$ being any external source.

We also consider the presence of a magnetic field, whose vector potential is $\mathbf{A}=(A_x,A_y,0)$, thus the Schr{\"o}dinger equation becomes
\begin{multline}
    \frac{1}{2}\left[\frac{(-i\hbar\partial_x +e A_x)^2}{m_x}+\frac{(-i\hbar\partial_y +e A_y)^2}{m_y} \right]\psi \\+\left(U_\mathrm{ext}-e\Phi\right)\psi=i\hbar \frac{\partial \psi}{\partial t},
\label{eq:schrodinger_mc}
\end{multline}
with $U_\mathrm{ext}$ being the total external potential, including the effect of the electron degeneracy. 

As was done before, we use the polar representation of the wavefunction~\eqref{eq:psi_madelung}, now defining the velocity field as:
\begin{eqnarray}
u_j=\frac{1}{m_j} \left(\hbar\partial_j\beta- qA_j\right),
\end{eqnarray}
which includes the vector potential, and eventually we arrive at the Euler equation
 \begin{multline}
\partial_t u_k+ 
\sum_j u_j\partial_j u_k 
+\frac{e}{ m_k}\sum_j u_j \left( \partial_kA_j- \partial_jA_k \right)
\\ = -\frac{\partial_k}{m_k}\left[
V^B+U_\mathrm{ext}-e(\Phi+\partial_tA_k)
\right]. \label{eq:Euler}
\end{multline}
For the external potential, we consider the Fermi degeneracy pressure, given for noninteracting anisotropic fermions by
\begin{equation}
U_\mathrm{ext}=V^F=\frac{\hbar^2 \pi}{2\sqrt{m_xm_y}}n^2.
\end{equation}

\subsection{Anisotropic magneto-plasmons}

Our first example for the usage of the Madelung formalism is to study anisotropic magneto-plasmons. For this, we consider a constant magnetic field characterized by the vector potential $\mathbf{A}=(B_zy/2,-B_z x/2,0)$.
Considering the linearized version of Eq.~\eqref{eq:Euler}, i.e., introducing $n=n_0+n_1$ and $\mathbf{u}=\mathbf{u}_1$, with $n_0$ being the equilibrium electronic density and $n_1$ and $\mathbf{u}_1$ the first-order fluctuations/corrections, we have
\begin{multline}
M\partial_t \mathbf{u}_1
+e \mathbf{u}_1 \times\mathbf{B}
= -\frac{\hbar^2}{4n_0 } \nabla^2\mathbf{\nabla} n_1 \\
+\frac{\hbar^2 \pi n_0}{\sqrt{m_xm_y}}\mathbf{\nabla} n_1-e\mathbf{\nabla}\Phi
, \label{eq:Euler2}
\end{multline}
with $M\equiv \mathrm{diag}(m_x,m_y)$ and $\mathbf{B}=(0,0,B_z)$. Likewise, the linearized version of the continuity equation becomes
\begin{equation}
\partial_t n_1 +n_0\mathbf{\nabla}_\parallel \cdot\mathbf{u}_1=0. \label{eq:cont_lin}
\end{equation}
We are searching for self-sustained solutions when the only source of electrostatic potential is the induced charge at the 2D material itself. Taking the Fourier transform, we can analytically obtain the solution of Eq.~\eqref{eq:Poisson}:
\begin{equation}
    \Phi(q)= \frac{e}{2\epsilon_0 q}n_1,
\end{equation}
with $q=\sqrt{q_x^2+q_y^2}$ denoting the in-plane wavenumber. Substituting back into the Fourier transform of Eqs.~\eqref{eq:Euler2} and \eqref{eq:cont_lin}, we arrive at a homogeneous system of equations:
\begin{subequations}
\begin{multline}
0=\left[\omega-\frac{q_x^2 }{ m_x\omega }\left(K_\mathbf{q}+V^{F}+V^C_\mathbf{q}\right)\right] u_{1,x}
\\ +\left[\frac{ie B_z}{ m_x }
-\frac{q_xq_y}{ m_x\omega }\left(K_\mathbf{q}+V^F+V^C_\mathbf{q}\right) 
 \right] u_{1,y}, 
\end{multline}
\begin{multline}
0=\left[-\frac{ie B_z}{ m_y}-\frac{q_xq_y }{ m_y\omega }\left(K_\mathbf{q}+V^F+V^C_\mathbf{q}\right)\right] u_{1,x} \\+\left[\omega-\frac{q_y^2 }{m_y\omega }\left(K_\mathbf{q}+V^F+V^C_\mathbf{q}\right)\right]u_{2,y}.
\end{multline} 
\label{eq:magn_pl}
\end{subequations}
Here, we have conveniently introduced the following energy contributions 
\begin{subequations}
\begin{align}
    K_\mathbf{q} &=
    E_0 \left[\frac{(q_x\ell_0)^2} {\mu_x}+\frac{(q_y\ell_0)^2 }{\mu_y}\right],\\
    V^{F} &= 
   E_0 \frac{2 \pi}{\sqrt{\mu_x \mu_y}},\\
     V_\mathbf{q}^{C} &=
     \frac{V_0}{q\ell_0},
\end{align}
\end{subequations}
where $K_\mathbf{q}$ is the free-electron kinetic energy with a characteristic Fermi degeneracy energy scale $E_0=\hbar^2/(2m_0\ell_0^2)=\hbar^2n_0/(2m_0)$, where $\ell_0=\sqrt{1/n_0}$ quantifies the average electron-electron distance in the Fermi sea and comes from the Bohm potential [Eq.~\eqref{eq:Bohm-potential}], $V^{F}$ is the energy associated with the Fermi pressure, and $V_\mathbf{q}^{C}$ is the Coulomb energy with a characteristic electrostatic energy scale $V_0=e^2\sqrt{n_0}/(2\epsilon_0)=e^2/(2\epsilon_0\ell_0)$. Finally, we have also introduced the relative effective masses $\mu_i=m_i/m_0$ (not to be confused with the vacuum permeability $\mu_0$).
We emphasize that $K_\mathbf{q}+V^B$ includes all the nonlocal corrections to the magnetoplasmon dispersion.

For a nontrivial solution of Eq.~\eqref{eq:magn_pl}, the determinant must vanish, yielding the plasmon dispersion relation
\begin{equation}
\hbar\omega_\mathbf{q}= \sqrt{
\left(K_\mathbf{q}+V^F+V_\mathbf{q}^C \right)K_\mathbf{q}
+(\hbar\omega_c)^2} \label{eq:magneto_plasmon_disp},
\end{equation}
where $\omega_c=eB_z/\sqrt{m_xm_y}$ is the cyclotron frequency. The above analysis shows how the problem is governed by three energy scales, $V_0$, $E_0$, and $\hbar\omega_c$ (the latter two depending on the anisotropic effective mass), while the spatial dispersion -- the dependence on wave number $q\ell_0$ -- is entirely scaled by the electron-electron distance $\ell_0$.

It is instructive that $\left(K_\mathbf{q}+V^B+V_\mathbf{q}^C \right)K_\mathbf{q}$ can be expanded to the desired order in $q\ell_0$ giving both linear [$V_\mathbf{q}^C K_\mathbf{q} \propto \vert q_i\ell_0\vert $], quadratic [$V^B K_\mathbf{q} \propto (q_i\ell_0)^2 $], and quartic [$ K_\mathbf{q}^2 \propto (q_i\ell_0)^4 $] contributions. As an example, for an isotropic electron gas ($\mu_x=\mu_y=1$) in the low-energy--low-wavenumber regime, we have
\begin{equation}
\hbar\omega_\mathbf{q}\simeq \sqrt{ V_0 E_0 \vert q\ell_0\vert
+(\hbar\omega_c)^2},\quad q\ell_0\ll 1 \label{eq:magneto_plasmon_disp-isotropic},
\end{equation}
exhibiting the well-known square-root dependence for magnetoplasmons in a 2DEG~\cite{Fetter:1985,Fetter:1986}. 

As we decrease the density (increasing $\ell_0$), the wavenumber where quantum effects become relevant also decreases. This term is also direction-dependent; therefore, in highly anisotropic systems, one direction can be significantly influenced by the quantum term, while the other remains essentially unaffected.

From the analysis of Eq.~\eqref{eq:magneto_plasmon_disp}, the term inside the first bracket contains the two nonlocal contributions $K_\mathbf{q}$ and $V_F$, originating from the Bohm potential and Fermi pressure, respectively, which can be compared with the Coulomb energy $V_\mathbf{q}^C$ to reveal the wavenumber scales governing each effect.

The wavelength scale that makes the Fermi-degeneracy term relevant, obtained by choosing $V^F/V_\mathbf{q}^C\sim 1$, is given by
\begin{equation}
    \lambda_\mathrm{FP} \sim \frac{a_0}{\sqrt{\mu_x\mu_y}}, \label{eq:fermi_wavelength}
\end{equation}
here intuitively parametrized in terms of the Bohr radius $a_0=4\pi\epsilon_0\hbar^2/m_0 e^2$. This result is independent of the electronic density. For given effective masses $\mu_x$ and $\mu_y$, we only need to consider the nonlocal correction due to the Fermi pressure when considering wavelengths smaller than $\lambda_\mathrm{FP}$. This regime is attained for in-plane hyperbolic plasmons, which can achieve very high wavenumbers for a given frequency. 

For the case of the quantum corrections due to the Bohm potential, setting $K_\mathbf{q}/V_\mathbf{q}^C\sim 1$, we obtain the wavelength
\begin{equation}
    \lambda_{B,j} \sim \sqrt[3]{ \mu_j a_0 \ell_0^2}, \label{eq:Bohm}
\end{equation}
which, due to the anisotropy, is direction-dependent and scales as $\ell_0^{2/3}$.

\subsection{Nonlocal anisotropic optical conductivity} \label{subsec:nonlocal_cond}

To obtain the linearized optical conductivity, we write Eq.~\eqref{eq:Euler2} with the scalar potential satisfying the Poisson equation and with an external contribution from an incident electric field $\mathbf{E}=\mathbf{E}^0e^{i(\mathbf{q}\cdot\mathbf{r}_\parallel-\omega t)}$ characterized by in-plane wavenumber $\mathbf{q}$ and frequency $\omega$,
\begin{multline}
    -e\Phi= \left(\frac{ eE^0_x}{iq_x}+ \frac{ eE^0_y}{iq_y}\right)e^{i(\mathbf{q}\cdot\mathbf{r}_\parallel-\omega t)}\\+\frac{\hbar^2\pi}{2\sqrt{m_xm_y}}n^2+\frac{e^2}{2\epsilon_0 q}n.
\end{multline}
In the spirit of the external field, we introduce the \emph{Ansatz} $\mathbf{u}=\mathbf{u}_1e^{i(\mathbf{q}\cdot\mathbf{r}_\parallel-\omega t)}$ and $n=n_0+n_1e^{i(\mathbf{q}\cdot\mathbf{r}_\parallel-\omega t)}$. With this, linearization of Eq.~\eqref{eq:Euler2} yields
\begin{multline}
M\partial_t \mathbf{u} +e\mathbf{u}_1\times\mathbf{B}
= \frac{\hbar^2}{4}\sum_j \frac{1}{m_j } \frac{\partial_j^2\mathbf{\nabla} n }{n_0}\\-e\mathbf{E}-\frac{\hbar^2 \pi}{ \sqrt{m_xm_y}} \mathbf{\nabla} n, \label{eq:lin}
\end{multline}
while the linearized continuity equation is still given by Eq.~\eqref{eq:cont_lin}. Substituting Eq.~\eqref{eq:cont_lin} in Eq.~\eqref{eq:lin}, and retaining terms to first order in $\mathbf{u}$, we obtain:
\begin{multline} 
M\omega \mathbf{u}_1
-\mathbf{q}\left(\frac{\hbar^2}{2}\sum_j \frac{q_j^2 }{m_j}+\frac{\hbar^2 \pi n_0}{ \sqrt{m_xm_y}} \right) \frac{\mathbf{q}\cdot\mathbf{u}_1}{\omega} \\ +ie \mathbf{u}_1\times\mathbf{B}
=-ie \mathbf{E}^0.
\label{eq:first_order_u}
\end{multline}
The solution is given by
\begin{equation}
\begin{pmatrix}
    u_{1,x} \\ u_{1,y}
\end{pmatrix} =\frac{-ie}{D(\mathbf{q},\omega) }\begin{pmatrix}
\omega -\frac{q_y^2f_\mathbf{q} }{\omega m_y}& \frac{g_\mathbf{q}(\omega)}{m_x}\\
  \frac{g_\mathbf{q}(\omega)}{m_y} & \omega -\frac{q_x^2f_\mathbf{q}}{\omega m_x}
\end{pmatrix} 
\begin{pmatrix}
\frac{ E_x}{m_x} \\ \frac{ E_y}{m_y}
\end{pmatrix},
\end{equation}
with 
\begin{equation}
D(\mathbf{q},\omega)=\omega^2-f_\mathbf{q}\left(\frac{q_x^2}{m_x}+ \frac{q_y^2}{m_y}\right)+\omega_c^2,
\end{equation}
and where we have introduced the functions
\begin{subequations}
\begin{multline}
    f_\mathbf{q}=E_0\big[ \mu_x^{-1}(q_x\ell_0)^2+\mu_y^{-1}(q_y\ell_0)^2\\+2\pi\mu_x^{-1/2}\mu_y^{-1/2}\big],
\end{multline}
\begin{equation}
g_\mathbf{q}=\frac{q_xq_y }{\omega } f_\mathbf{q}-i eB_z. 
\end{equation}
\end{subequations}

The first-order current is $\mathbf{J}_1=-en_0\mathbf{u}_1$. Using the constitutive equation $\mathbf{J}=\sigma \mathbf{E}$, we obtain the hydrodynamic conductivity tensor
\begin{multline}
\sigma(\mathbf{q},\omega)=\frac{\sigma_0(\omega) }{\omega^2-\omega^2_\mathbf{q} }\frac{m_0}{m_x m_y} \\ \times \begin{pmatrix}
 m_y\omega^2 - q_y^2 f_\mathbf{q} & -ieB_z\omega+ q_xq_y f_\mathbf{q}\\
 i e B_z\omega+ q_xq_y f_\mathbf{q} & m_x\omega^2 -q_x^2f_\mathbf{q}
\end{pmatrix}, \label{eq:cond_fin}
\end{multline}
where we have introduced $\sigma_0(\omega)=in_0e^2 m_0^{-1}\omega^{-1}$
and $\omega^2_\mathbf{q}=(q_x^2/m_x+q_y^2/m_y)f_\mathbf{q}$. 

Equation~\eqref{eq:cond_fin} represents the nonlocal optical conductivity of an electron gas with anisotropic mass. As discussed in the previous section, this equation accounts for two distinct sources of nonlocality: Fermi pressure and the Bohm potential. We emphasize that the local Drude expression rigorously follows from Eq.~\eqref{eq:cond_fin} in the limit $q\rightarrow 0$.

\section{Application to phosphorene and black phosphorus} \label{sec:applications}
 
Black phosphorus is a van~der~Waals material with a layered orthorhombic crystal structure~\cite{Xia2019}. Its phosphorus (P) atoms form a puckered honeycomb lattice, allowing us to define two orthogonal crystalline directions: armchair (AC) and zigzag~(ZZ). As a semiconductor, its bandgap varies with the number of layers, ranging from 0.3\,eV in the bulk to 2.0\,eV in the monolayer~\cite{Tran2014}. The mechanical and optical properties of black phosphorus reflect its intrinsic anisotropy~\cite{Wang2016}. In this section, we begin with the monolayer case -- phosphorene~\cite{Liu:2014} -- before exploring surface polaritons in few-layer black phosphorus.

\subsection{\emph{Ab~initio} calculations} \label{subsec:abinitio}

\begin{figure}
    \centering
    \includegraphics[width=1\columnwidth]{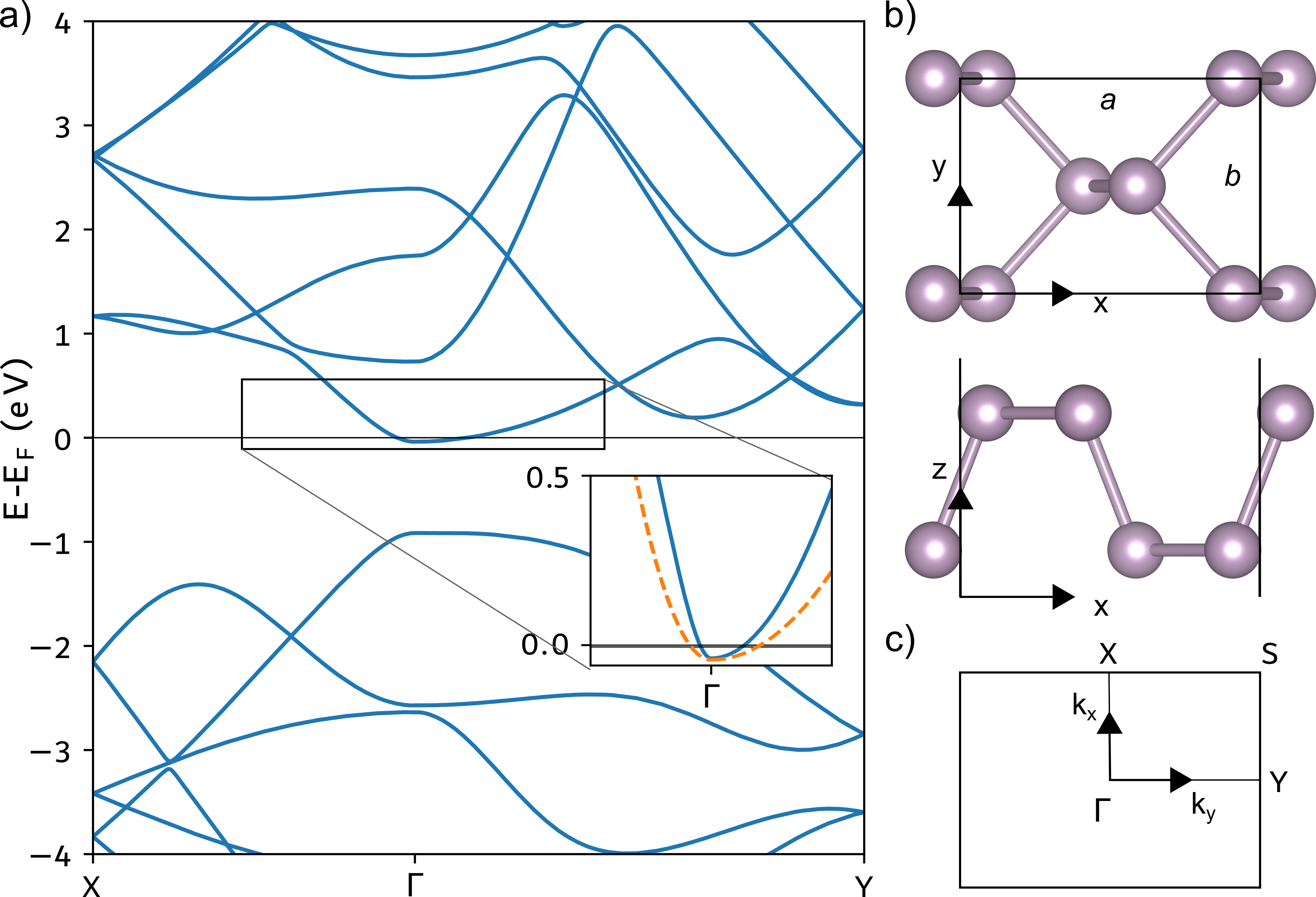}
    \caption{Electronic energy band structure of phosphorene. Panel (a) shows the band structure of extended monolayer phosphorene with doping charge carrier density $n=1\times10^{13}$\,cm$^{-2}$. The inset shows the conduction band along the $\mathrm{\Gamma}$--X and $\mathrm{\Gamma}$--Y directions of the first Brillouin zone (BZ) together with the parabolic bands (orange dashed lines) resulting from the fitted effective masses extracted from Fig.~\ref{fig:ab_initio}. Panel (b) shows the unit cell of the 2D phosphorene crystal seen from above (top) and the side (bottom). Panel (c) shows the first BZ of the 2D crystal.}
    \label{fig:bandstructure}
\end{figure}

\begin{figure*}[ht!]
\centering\includegraphics[width=0.9\linewidth]{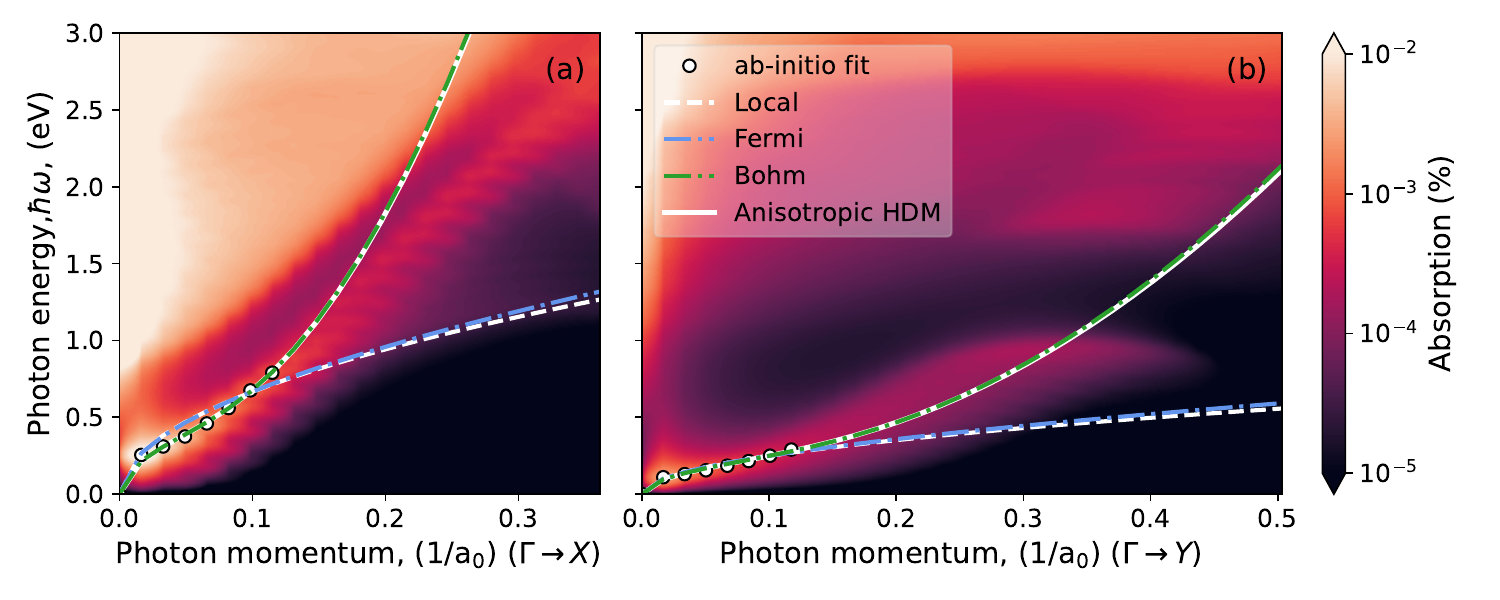}
    \caption{Density plot of the absorption obtained from \emph{ab~initio} calculations as a function of the the photon energy $\hbar\omega$ and the photon wavenumber $q$ with zero magnetic field. Panel (a) shows variations of $q$ along the $\mathrm{\Gamma}$--X direction in the Brillouin zone of the electronic-structure problem, while panel (b) is for the $\mathrm{\Gamma}$--Y direction. In both panels, results for the dispersion relation from the hydrodynamic model [Eq.~\eqref{eq:magneto_plasmon_disp}] are superimposed, showing both the local approximation (dashed white line), the additional effect of Bohm's potential (dash-dotted green line), the Fermi pressure (dash-dotted blue line), and the combined effects of Bohm's potential and the Fermi Pressure (solid white line). Maxima in absorption from the first principles calculation at low momenta are used for fitting of the effective masses (circles).
    }
    \label{fig:ab_initio}
\end{figure*}

We model an extended monolayer of phosphorene using DFT and the Perdew--Burke--Ernzerhof (PBE) exchange-correlation (XC) functional~\cite{PBE} as implemented in the Quantum Espresso (QE) software~\cite{QE}. The Kohn--Sham wavefunctions are expanded in a plane-wave basis with an energy cutoff of 48\,Rydberg (Ry) and using norm-conserving pseudo potentials from the Pseudo Dojo database~\cite{hamann2013optimized,van2018pseudodojo}. We include 40 bands per unit cell, the reciprocal space is resolved by a $k$-point grid with 8 points per {\AA}ngstr{\"o}m ({\AA}), and Gaussian smearing of 0.001\,Ry is used for describing the occupations. Relaxing the atom positions until a force tolerance of 0.02\,eV/{\AA}$^3$ results in lattice constants of $a=4.61$\,{\AA} and $b=3.34$\,{\AA}, while 1.5\,nm of vacuum separation is included between repeating images in the $z$-direction. The unit cell of the relaxed phosphorene crystal can be seen in Fig.~\ref{fig:bandstructure}\pnl{b}. Electrostatic doping is included by adding an electron charge carrier density of $10^{13}$\,cm$^{-2}$, resulting in a Fermi level crossing the bottom of the conduction band as seen in the resulting band structure shown in Fig.~\ref{fig:bandstructure}\pnl{a}. The absorption spectra presented in Fig.~\ref{fig:ab_initio} are calculated within the RPA using the Yambo software~\cite{marini2009yambo,sangalli2019manybody}, including all the Kohn--Sham bands and a 4 times denser $k$-point sampling. We use 2D truncated Coulomb interaction~\cite{PhysRevB.96.075448} both for the QE and Yambo calculations to avoid interactions between repeating images.

Ground-state DFT using the PBE functional is known to significantly underestimate the band gap of 2D materials, as the method excludes the many-body effects which result in a significantly reduced screening. In these calculations, we obtain a band gap of 0.89\,eV, which should be compared to experimental measurements yielding a gap size of 2\,eV~\cite{Liang:2014}. However, since many-body calculations using the GW method mainly result in a nearly constant energy correction of the conduction band~\cite{PhysRevB.96.115431}, the effective masses and plasmon dispersion are not expected to differ much from what we extract within the single-particle approximation. 

\subsection{Comparison of the First Principles and hydrodynamic approaches} \label{subsec:comparison}

Figure~\ref{fig:ab_initio} presents the \emph{ab~initio} results for plasmons in phosphorene (colormap), as described in the previous section, considering $n_0=10^{13}$\,cm$^{-2}$. Additionally, we show the analytical results derived from Eq.~\eqref{eq:magneto_plasmon_disp} for a zero magnetic field ($B_z=0$), incorporating different nonlocal terms of $f_\mathbf{q}$. The effective masses $m_x$ and $m_y$ were determined by fitting the plasmon dispersion at low momenta in both the zigzag and armchair directions, yielding $m_x = 0.464m_0$, $m_y = 2.394m_0$. The parabolic bands resulting from these effective masses are shown by the dashed orange lines on the inset of Fig.~\ref{fig:bandstructure}\pnl{a}. It can be seen that the conduction band along the $\mathrm{\Gamma}$--X direction is almost linear close to the $\mathrm{\Gamma}$-point, resulting in a poorer fit to the parabolic band along this direction. Including the Bohm potential, a nonlocal contribution, enables an accurate description of the \emph{ab~initio} plasmon dispersion, whereas the Drude term, which follows a square root dependence on the wavenumber, fails to do so. Additionally, for the electronic density considered, the contribution from Fermi degeneracy pressure is negligible.

\begin{figure*}
    \centering
    \includegraphics[width=\linewidth]{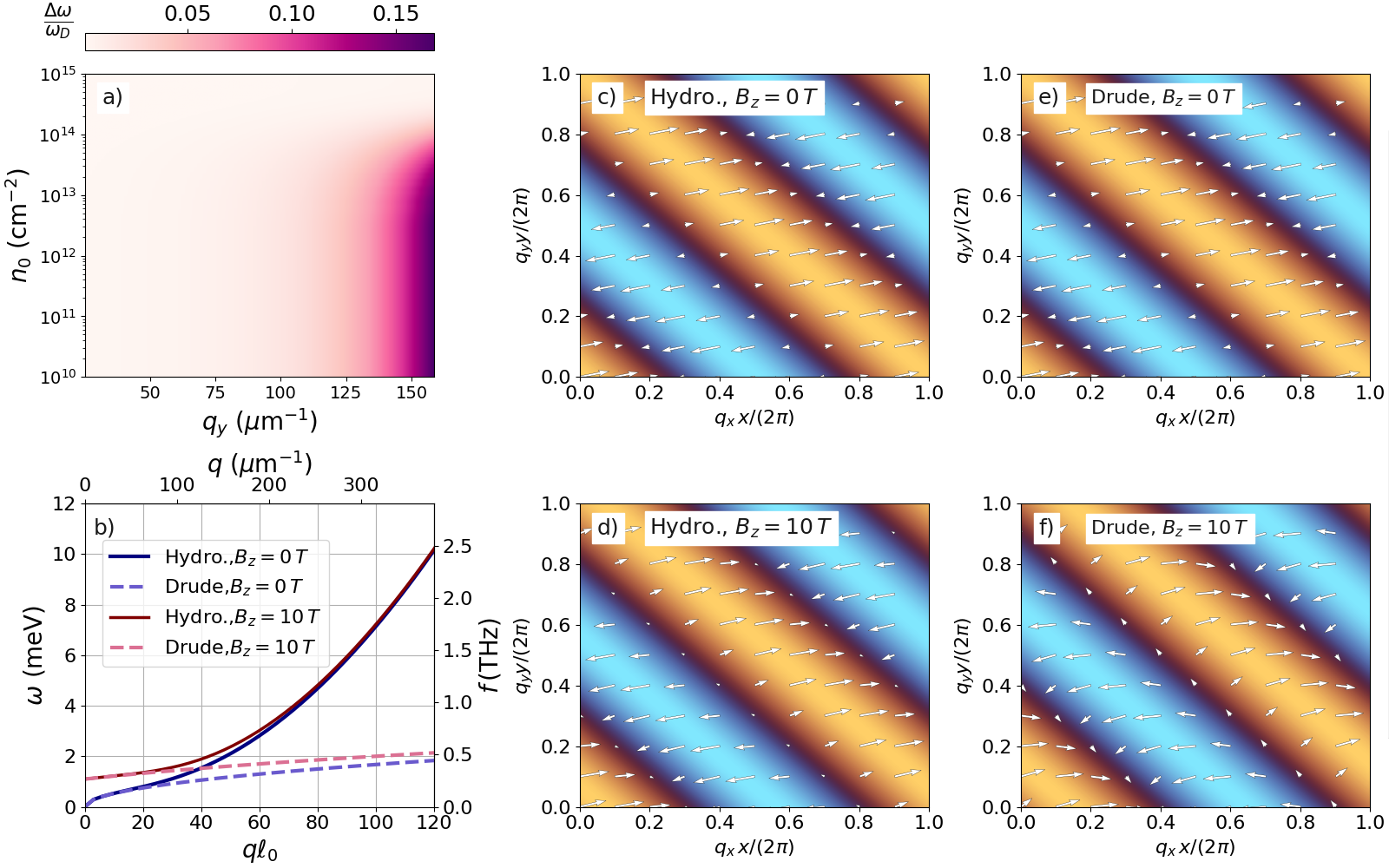}
    \caption{Panel (a) is a density plot of $\Delta\omega/\omega_0$ -- the relative difference between the nonlocal and local plasmon dispersion of phosphorene -- as function of the electronic density $n_0$ and the plasmon wavenumber $q_y$ in the zigzag direction. 
    Panel (b) shows the magnetoplasmon dispersion relation for phosphorene -- frequency $f=\omega/2\pi$ versus wavenumber $q$ -- comparing the nonlocal hydrodynamic model (solid lines) to the local Drude model (dashed lines) for both $B_z=0$\,T (blue coloring) and $B_z=10$\,T (red coloring). The effective mass was obtained from the \textit{ab~initio} fitting and $n_0=10^9$\,cm$^{-2}$. We consider a direction that makes a 45$^\circ$ angle with either the zigzag or armchair direction. Panels (c-f) show Quiver plots of the plasmon velocity field and colormap of the charge-density fluctuation for phosphorene for $q\ell_0=90$, as a function of the real-space coordinates $x$ and $y$ for different magnetic fields, for both the nonlocal hydrodynamic and the local Drude models. We use the effective mass obtained from the \emph{ab~initio} fitting and $n_0=10^9$\,cm$^{-2}$.}
    \label{fig:magneto_plasmons}
\end{figure*}

\subsection{Magnetoplasmons} \label{subsec:magnetoplasmons}

We begin this section by examining the nonlocal effects on plasmon dispersion governed by Eq.~\eqref{eq:magn_pl}, using the fitted effective masses from the previous section. In Fig.~\ref{fig:magneto_plasmons}\pnl{a}, we present a colormap of the relative difference in plasmon dispersion, $\Delta \omega/\omega_D$, where $\omega_D$ represents the Drude dispersion, obtained by setting $f_\mathbf{q}=0$ in Eq.~\eqref{eq:magn_pl}. The quantity $\Delta \omega\equiv \omega_p-\omega_D$ is shown as a function of the electronic density $n_0$ and the wavenumber $q_y$.
We observe that nonlocal effects are present in the low electronic density and high wavenumber region, such as that for $n_0< 10^{15}$\,cm$^{-2}$ and $q_y>60\,\upmu$m$^{-1}$, the nonlocal corrections are not negligible. A similar conclusion could be obtained for the $q_x$ dependence, but nonlocal effects become important at smaller $q_x$ when compared to $q_y$ due to $m_x<m_y$.

The magnetoplasmon spectrum derived from Eq.~\eqref{eq:magneto_plasmon_disp} is displayed in Fig.~\ref{fig:magneto_plasmons}\pnl{b}. In the presence of a magnetic field, we observe that the deviation of the hydrodynamic (solid curves) and Drude model (dashed curves) shifts toward higher wavenumbers. The behavior is different for the plasmon velocity fields. In Fig.~\ref{fig:magneto_plasmons}\pnl{c-f}, we present a quiver plot of the normalized velocity fields obtained from the nontrivial solutions of Eqs.~\eqref{eq:magn_pl}. The colormap represents the normalized charge density calculated from the continuity equation~\eqref{eq:cont_lin} for both finite and zero magnetic fields, considering the nonlocal hydrodynamic and local Drude ($f_\mathbf{q}=0$) models.

In the case of a vanishing magnetic field, considered in Figs.~\ref{fig:magneto_plasmons}\pnl{c} and \ref{fig:magneto_plasmons}\pnl{e}, we can show that the nontrivial solution of Eqs.~\eqref{eq:magn_pl} simplifies to $q_yu_{1,x}/m_y-q_xu_{1,y}/m_x=0$, and does not depend on $f_\mathbf{q}$, such that the velocity field $(u_{1,x},u_{1,y})$ is parallel to the vector $(q_x/m_x,q_y/m_y)$ in both cases: Drude and hydrodynamic models. In this case, the velocity field oscillates perpendicular to the charge density wavefront. However, for finite $B_z$, the velocity field acquires a dependence on nonlocal corrections, as can be seen by comparing Figs.~\ref{fig:magneto_plasmons}\pnl{d} and \ref{fig:magneto_plasmons}\pnl{f}. 
In the absence of nonlocal corrections, as shown in Fig.~\ref{fig:magneto_plasmons}\pnl{f}, the presence of the magnetic field causes the velocity field to become oblique to the charge density wavefront.
In the nonlocal hydrodynamic model, as illustrated in Fig.~\ref{fig:magneto_plasmons}\pnl{d}, nonlocal effects decrease the transverse component of the velocity field, thus diminishing the magnetic field effect.

\subsection{Effects of nonlocality: polaritonic spectrum} \label{subsec:spp}

\begin{figure}
    \centering
    \includegraphics[width=0.9\linewidth]{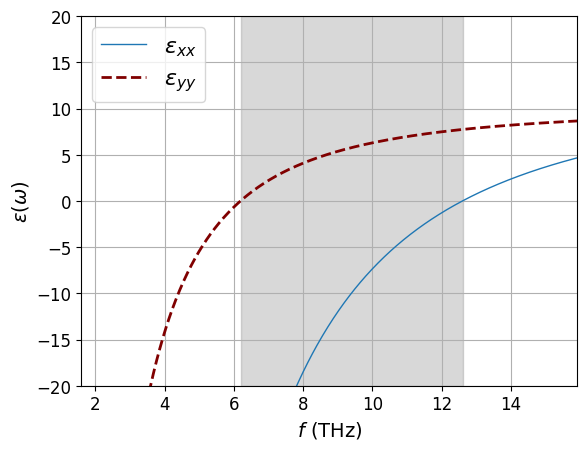}
    \caption{Black phosphorus anisotropic dielectric function $\epsilon(\omega)$, calculated from Eq.~\eqref{eq:diel} for an electronic density $n_0=10^{12}$\,cm$^{-2}$. The Reststrahlen band with $\Re\{\epsilon_{xx}\} \times \Re\{\epsilon_{yy}\}<0$ is highlighted in grey.}
    \label{fig:dielectric_function}
\end{figure}

\begin{figure*}
    \centering
    \includegraphics[width=\linewidth]{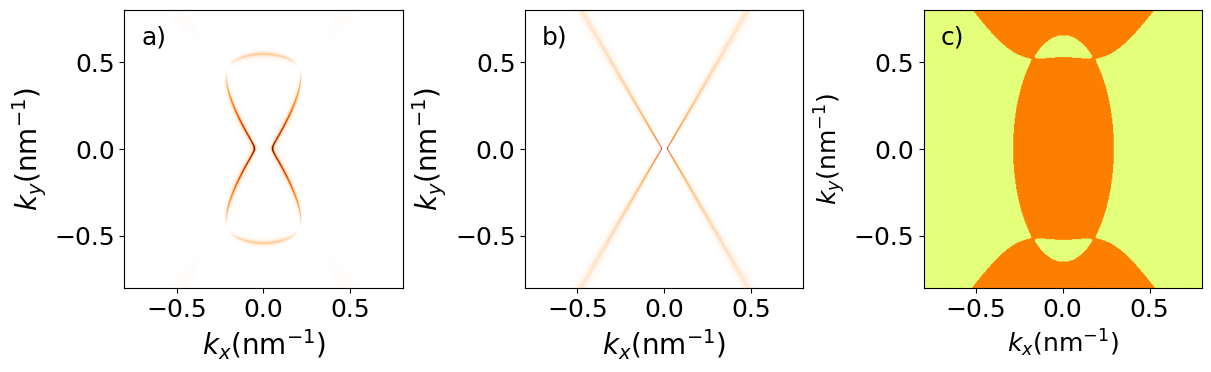}
    \caption{Panels (a) and (b) show density plots of the loss function $\Im\{r_{pp}(\mathbf{q},\omega)\}$ as a function of $\mathbf{q}$ for a BP thin film at $f=\omega/2\pi=9.0$\,THz calculated using Eq.~\eqref{eq:r_pp}. Panel (a) is for the nonlocal hydrodynamic model, while panel (b) is for the local Drude model. Panel (c) is a zero-contour plot of $\Im\{\lambda_1\lambda_2\}$ in wavenumber space, separating positive regimes (in yellow) from negative regimes (in orange). The plot is for a photon frequency $f=9.0$\,THz and the eigenvalues $\lambda_{1,2}$ are those of the optical conductivity tensor matrix [Eq.~\eqref{eq:total_cond}].}
    \label{fig:loss_functions}
\end{figure*}

The coupling of light with plasmons gives rise to surface plasmon-polaritons (SPPs). In the case of BP, the anisotropic nature of the interband response indicates the existence of in-plane hyperbolic SPPs~\cite{biswas2021tunable}. In this section, we examine a slab of BP with negligible thickness. The total optical conductivity of BP is expressed as the sum of a hydrodynamic component given by Eq.~\eqref{eq:cond_fin} and an additional contribution from interband transitions that is not accounted for in the hydrodynamic model,
\begin{equation}
\sigma_{ij}(\mathbf{q},\omega)=\sigma_{ij}^\mathrm{hydro}(\mathbf{q},\omega)+\sigma_{i}^\mathrm{inter}(\omega)\delta_{ij}. \label{eq:total_cond}
\end{equation}
Here, the interband conductivity $\sigma_{i}^\mathrm{inter}(\omega)=i\epsilon_0 \epsilon_i t$ is given in terms of constant relative permittivities $\epsilon_i$ and the BP slab thickness $t$, which we take from data extracted from IR absorption experiments~\cite{biswas2021tunable}: $\epsilon_{x}=12.5$, $\epsilon_{y}=10.2$, and $t=2.9$\,nm. In Fig.~\ref{fig:dielectric_function} we show the dielectric function 
\begin{equation}
\epsilon_{ii}(\omega)= -i\frac{\sigma_{ii}(\mathbf{q}=0,\omega)}{\epsilon_0 t}, \label{eq:diel}
\end{equation} 
for both armchair and zigzag directions of BP, showing in the shaded region the BP Reststrahlen band, i.e., when $\Re\{\epsilon_{xx}\} \times \Re\{\epsilon_{yy}\}<0$. In this section, for the thin-layer BP, we will consider the effective masses of Ref.~\onlinecite{biswas2021tunable}, $\mu_x=0.14$ (AC) and $\mu_y=0.71$ (ZZ).

The presence of SPPs can be seen in the TM loss function, which is calculated from the imaginary part of the Fresnel coefficient $r_{p,p}$ [see App.~\ref{app:loss} for the derivation]:
\begin{multline}
r_{p,p}=\frac{1}{D^\prime}\Bigg[
\left( \frac{2k_z}{k_0} -\frac{k_z^2}{k_0^2} \mu_0c\tilde{\sigma}_{xx}\right)
\left(\frac{2k_z}{k_0} -\mu_0c\tilde{\sigma}_{yy} \right)
 \\ -\left(\frac{k_z}{k_0}\mu_0c\right)^2\tilde{\sigma}_{xy}\tilde{\sigma}_{yx}
-\frac{2k_z}{k_0}\left(\frac{2k_z}{k_0}-\mu_0c\tilde{\sigma}_{yy} \right)\Bigg], \label{eq:r_pp}
\end{multline}
where
\begin{multline}
D^\prime=\left( \frac{2k_z}{k_0} -\frac{k_z^2}{k_0^2} \mu_0c\tilde{\sigma}_{xx}\right)
\left(\frac{2k_z}{k_0} -\mu_0c\tilde{\sigma}_{yy} \right)
\\-\left(\frac{k_z}{k_0}\mu_0c\right)^2\tilde{\sigma}_{xy}\tilde{\sigma}_{yx}, \label{eq:rpp_den}
\end{multline}
with $\mu_0=c^{-2}\epsilon_0^{-1}$ being the vacuum permeability (not to be confused with relative effective masses $\mu_x$ and $\mu_y$), while $\tilde{\sigma}_{ij}$ are the components of the rotated conductivity tensor $\tilde{\sigma}=R^T_\phi\sigma R_\phi$. Here, $R_\phi$ is the 2D rotation matrix along the $z$ axis and $\tan \phi=q_x/q_y$.

In Fig.~\ref{fig:loss_functions}\pnl{a-b}, we show the TM loss function $ \Im \{r_{p,p}\}$ for the surface plasmon-polariton isofrequency for $f=9.0$\,THz, which lies within the Reststrahlen band when $n_0= 10^{12}$\,cm$^{-2}$. The results are shown for both the nonlocal hydrodynamic and local Drude models, including losses (see App.~\ref{app:losses}), where we fixed the damping rate at $\hbar\gamma=1.0$\,meV. 
The models exhibit markedly different qualitative behaviors. While the local Drude model predicts the existence of hyperbolic modes, these modes completely vanish when nonlocal effects are taken into account. Instead, the isofrequency transitions toward what is expected for a typical anisotropic surface plasmon-polariton. This behavior can be understood by analyzing Eq.~\eqref{eq:cond_fin}; the increase in wavenumber alters the sign of the imaginary part of the eigenvalues of the conductivity tensor due to its dependence on $q$.

In Fig.~\ref{fig:loss_functions}\pnl{c}, we illustrate the sign of the imaginary part of the product of the eigenvalues of the conductivity tensor~\eqref{eq:total_cond}. When the sign is negative, we can expect the presence of hyperbolic modes. Conversely, when the sign is positive, we may observe elliptic modes (if both imaginary parts are negative) or no modes at all (if both are positive). By comparing Figs.~\ref{fig:loss_functions}\pnl{b} and \ref{fig:loss_functions}\pnl{c}, we observe that in the hyperbolic branches of the Drude model, the hydrodynamic model alters the sign of $\Im\{\lambda_1\lambda_2\}$, thereby inhibiting the formation of hyperbolic modes.
Similar conclusions have been drawn for SPPs in BP using the Kubo formula for optical conductivity~\cite{Correas-Serrano_2016}, considering an effective Hamiltonian~\cite{Rodin2014}. It was observed that the inclusion of nonlocal effects significantly alters the isofrequencies, preventing the emergence of hyperbolic modes. A similar conclusion has been reached in the context of hyperbolic metamaterials~\cite{Mortensen2012} (and eventually also natural hyperbolic materials~\cite{Gjerding:2017}), where the incorporation of nonlocal effects strongly renormalizes the spectrum in the high wavenumber limit.

\subsection{Dipoles and Purcell Effect} \label{subsec:purcell}

One possible way to launch surface plasmon-polaritons is through the use of an emitting dipole~\cite{gonccalves2016introduction}, where the near-field of the dipole couples with the light-matter modes. In the scanning near-field optical microscopy (SNOM), the atomic-force microscopy (AFM) tip can be modeled as a point dipole~\cite{ Keilmann2004}; in this case, the tip supports evanescent modes that are excited by an incoming laser and can couple to SPP modes. To obtain the electromagnetic field in the presence of a dipole, we use the dyadic Green function formalism~\cite{novotny2012principles}, in which, for the case of light impinging on an anisotropic medium, it was obtained elsewhere~\cite{Gomez-Diaz2015}:
\begin{flalign}
    G(\mathbf{r},\mathbf{r}^\prime,\omega)=\frac{i}{8\pi^2} \,
    \int d\mathbf{k}_\parallel \sum_{jk=s,p} M_{jk}r_{jk} \nonumber \\ \times e^{ik_z|z+z^\prime|} e^{i \mathbf{k}_\parallel \cdot (\mathbf{r}_\parallel-\mathbf{r}_\parallel^\prime)}, \label{eq:dyadic}
\end{flalign}
where $r_{ss}$, $r_{sp}$, $r_{ps}$, and $r_{pp}$ are the Fresnel coefficients calculated in App.~\ref{app:loss}, and the matrix $M_{jk}$ is defined in App.~\ref{app:dya}. However, the numerical calculation of the integral above is challenging due to the presence of the SPP poles in the Fresnel coefficients. The electric field for an electric dipole that oscillates with frequency $\omega$ located at the position $\mathbf{r}^\prime$ is obtained from the dyadic Green's function as:
\begin{equation}
    \mathbf{E}(\mathbf{r},\omega)= \omega^2 \mu_0\left[ G_0(\mathbf{r},\mathbf{r}^\prime,\omega) +G(\mathbf{r},\mathbf{r}^\prime,\omega) \right]\cdot\mathbf{d},
\end{equation}
where $\mathbf{d}$ is the electric dipole moment and $G_0$ is the free-space dyadic Green's function.

\begin{figure*}[ht!]
\centering\includegraphics[width=0.9\linewidth]{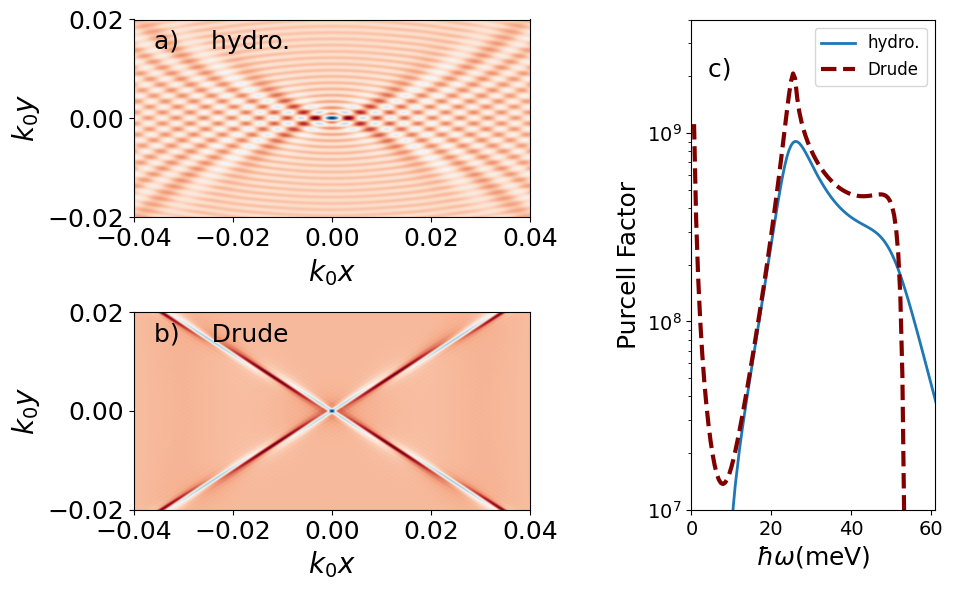}
    \caption{Panels (a) and (b) show plasmonics wakes induced by an electric dipole aligned with the $z$-axis with a frequency $\omega/(2\pi)=9$\,THz. The colormap shows the electric field in the $z$-direction at the surface of the BP. Panel (a) is the hydrodynamic model and panel (b) the Drude model. Panel (c) compares the Purcell factor computed from the hydrodynamic model (solid line) and Drude model (dashed line) as a function of the frequency of the electric dipole. In all the plots, the dipole-BP sheet distance is $z_0=3.0$\,nm and the BP electronic density is $n_0=10^{12}$\,cm$^{-2}$.}
    \label{fig:dipoles}
\end{figure*}

For simplicity, we henceforth consider an electric dipole oriented along the $z$-axis; accordingly, the reflected electric field in the $z$-direction is given by
\begin{multline}
    E_{\mathrm{ref},z}(\mathbf{r})=\frac{i\omega^2\mu_0d_z}{8\pi^2}
    \int d \mathbf{k}_\parallel \frac{k_\parallel^2}{k_0^2k_z}r_{pp} \\ \times e^{ik_z|z+z^\prime|} e^{i \mathbf{k}_\parallel \cdot (\mathbf{r}_\parallel-\mathbf{r}_\parallel^\prime)}. \label{eq:Ez}
\end{multline}
The integral is dominated by the plasmon-pole, which can be obtained from the Fresnel coefficient denominator~\eqref{eq:rpp_den} as the solution of
\begin{multline}
     \left( \frac{2k_z}{k_0} -\frac{k_z^2}{k_0^2} \mu_0c\tilde{\sigma}_{xx}\right)
\left(\frac{2k_z}{k_0} -\mu_0c\tilde{\sigma}_{yy} \right)
\\-\left(\frac{k_z}{k_0}\mu_0c\right)^2\tilde{\sigma}_{xy}\tilde{\sigma}_{yx}=0.
\end{multline}
For the case of plasmons, this pole is dominated by the TM part, i.e., the above equation has approximately the same roots as:
\begin{equation}
\frac{2k_z}{k_0} -\frac{k_z^2}{k_0^2} \mu_0c\tilde{\sigma}_{xx} =0, \label{eq:new_pole}
\end{equation}
and we approximate the $r_{pp}$ Fresnel coefficient as
\begin{equation}
    r_{pp}= \frac{F(\omega)}{\omega-\omega_\mathrm{SPP}(\mathbf{q})},
\end{equation}
with $\omega_\mathrm{SPP}(\mathbf{q})$ being the solution of Eq.~\eqref{eq:new_pole} and $F(\omega)$ corresponding to the residue of the plasmon-polariton pole, 
\begin{equation}
    F(\omega)=-2\omega.
\end{equation}

We next shift the pole by an infinitesimal amount and take the imaginary part:
\begin{equation}
    \Im\{r_{pp}(\mathbf{q},\omega)\}=-\pi F(\omega) \delta[\omega-\omega_\mathrm{SPP}(\mathbf{q})].
\end{equation}
Now to simplify the integral in Eq.~\eqref{eq:Ez}, we rewrite it in polar coordinates and make the change of variables $k_\parallel=k_0 s$. Considering $s\gg 1$, we obtain for a dipole placed at $(x^\prime,y^\prime,z^\prime)=(0,0,z_0)$ that
\begin{multline}
    E_z= \frac{k_0^3d_z}{4\pi\epsilon_0} \int ds s^2e^{-sk_0|z+z_0|}\int d\phi \,\delta(\omega-\omega_\mathrm{SPP}) \\ 
    \times\exp\left[{isk_0(x\cos\phi+y\sin\phi)
    } \right]. \label{eq:Ez2}
\end{multline}
To illustrate our point, we will for simplicity neglect the Bohm potential \eqref{eq:Bohm-potential}, so that we can obtain an analytical expression for the isofrequencies of the dispersion relation, as detailed in App.~\ref{app:spp_disp}. The relation between $s$ and the polar angle $\phi$, obtained after integrating over $\phi$ using the Dirac delta function, is
\begin{equation}
    \cos(2\phi_{\scriptscriptstyle H}) = -\frac{t_1}{2t_2}+\sqrt{\left(\frac{t_1}{4t_2}\right)^2-\frac{t_0}{t_2}},\label{eq:cos2phi}
\end{equation}
where we have defined
\begin{subequations}
\begin{equation}
    t_2=-C_p s^2 \frac{\Delta \mu}{\mu_y\mu_x}k_0d\frac{\Delta \epsilon}{4},
\end{equation}
\begin{multline}
    t_1=\frac{\Delta \mu}{2}A+ \frac{C_ps\Delta \mu}{2\mu_x\mu_y} \left[k_0d\left( \epsilon_m-1 \right)s+2\right] \\ 
    +\left(1-\frac{C_ps^2}{2\mu_r} \right)\frac{k_0d \Delta\epsilon}{2} , 
\end{multline}
\begin{multline}
    t_0=\frac{\mu_x+\mu_y}{2}A\\
    +\left(\frac{1}{s}-\frac{C_ps}{2\mu_r}\right) \left[k_0d\left(1-\epsilon_m \right)s-2 \right],
\end{multline}
\end{subequations}
and $C_p=\frac{\pi n_0 \tilde{\lambda}_c^2}{\sqrt{\mu_x\mu_y}}$, $A=\frac{4\pi \alpha \tilde{\lambda}_c n_0}{k_0\mu_x\mu_y}$, $\epsilon_m=\frac{\epsilon_x+\epsilon_y}{2}$, $\Delta\epsilon=\epsilon_y-\epsilon_x$, $\mu_\mathrm{r}^{-1}=\mu_x^{-1}+\mu_y^{-1}$, $\Delta \mu=\mu_y-\mu_x$, with $\alpha$ the fine structure constant and $\tilde{\lambda}_c$ the reduced Compton wavelength. 
Based on this, the electric field in Eq. (\ref{eq:Ez2}) can be express -- after integrating in $\phi$ using Dirac's delta function -- as a single quadrature that can be easily computed numerically:
\begin{multline}
E_z = 
 \frac{k_0^3d_z}{4\pi\epsilon_0} \\\times \int ds s^2e^{-sk_0|z+z^\prime|}\sum_{\phi_{\scriptscriptstyle H}} \,\left|\frac{\partial }{\partial \phi}\omega_\mathrm{SPP}(s,\phi_{\scriptscriptstyle H}) \right|^{-1} \\ 
    \times\exp\left[{i sk_0(x\cos\phi_{\scriptscriptstyle H}+y\sin\phi_{\scriptscriptstyle H})
    } \right],
\end{multline}
where the summation in $\phi_{\scriptscriptstyle H}$ corresponds to all nondegenerate solutions of Eq.~\eqref{eq:cos2phi} and the integral is limited to the intervals where $|\cos2\phi_{\scriptscriptstyle H}(s)|\le1$. Using the symmetry of the solutions, we can write the above integral as:
\begin{multline}
    E_z(\mathbf{r})= \frac{k_0^3d_z}{4\pi\epsilon_0} \int ds \,\frac{e^{-sk_0|z+z_0|}}{|D^{\prime\prime}|}\\ 
    \times \cos(sk_0 x \cos \phi_{\scriptscriptstyle H} )\cos(sk_0y\sin \phi_{\scriptscriptstyle H} ),
\end{multline}
with
\begin{multline}
    D^{\prime\prime}=\frac{A(\mu_x-\mu_y)}{\left(1-C_ps^2u_\phi\right)^2}
    \Big[ 1 - C_p s^2 (u_\phi-1) \\ +k_0 d (\epsilon_y - \epsilon_x) \Big] \sin(2\phi_{\scriptscriptstyle H}). 
\end{multline}

For the case of the local response Drude model, Eq.~\eqref{eq:cos2phi} simplifies to:
\begin{equation}
    \cos(2\phi_{\scriptscriptstyle D})= \frac{4-s(\mu_x+\mu_y)A- sk_0d[2-(\epsilon_x+\epsilon_y)]}{s(\mu_y-\mu_x)A+sk_0d(\epsilon_y-\epsilon_x)}.
\end{equation}

In Figs.~\ref{fig:dipoles}\pnl{a} and \ref{fig:dipoles}\pnl{b}, we show the comparison between the hydrodynamic (including only the Fermi pressure term) and Drude models, for the frequency $\omega/(2\pi)=9$\,THz, that lies inside the Reststrahlen band. While the Drude model predicts extremely canalized plasmons that propagate in the direction dictated by the hyperbolic dispersion, the hydrodynamic model shows closed wavefronts propagating in all directions.

The presence of a 2D material that supports plasmons changes a quantum emitter radiative decay through the Purcell effect~\cite{Ferreira2020}. The study of the Purcell effect in phosphorene, including strain, was done before~\cite{Abrantes2023} but neglecting nonlocal effects, i.e., the phosphorene optical conductivity was calculated using a tight-binding model without dependence on the in-plane wavenumber, or, when including nonlocal effects, the anisotropy was averaged~\cite{Petersen2017}. The Purcell factor for a dipole aligned in the $z$-direction can be calculated using the following equation~\cite{Abrantes2023}:
\begin{equation}
\frac{\Gamma}{\Gamma_0}= 1 +\frac{3}{4\pi k_0^3} \Im\left[\int d\mathbf{k}_\parallel \frac{k_\parallel^2e^{2i\sqrt{k_0^2-k_\parallel^2}z_0} }{\sqrt{k_0^2-k_\parallel^2}} r_{pp}\right]. \label{eq:purcell}
\end{equation}
The details of the numerical calculation of the Purcell factor are given in App.~\ref{app:Purcell} and the results are shown in Fig.~\ref{fig:dipoles}\pnl{c}. While the Drude model predicts a peak in the zero frequency, this peak disappears in the hydrodynamic model. The peaks of the Drude model also decrease by almost a factor of 3. For higher frequencies, the hydrodynamic model predicts a higher Purcell factor.

\section{Conclusions and Perspectives} \label{sec:ending}

Through various examples, we have unveiled the critical importance of nonlocal electrodynamics in addressing anisotropic materials, particularly those exhibiting hyperbolic behavior. The formalism presented here can be effectively employed to calculate higher-order responses in such materials, thus reaching the nonlinear regime. Furthermore, it offers a robust framework for describing nanostructures incorporating 2D materials, such as quantum dots or arrays of nanoribbons.

Recent advances in fabrication techniques have yielded systems with reduced losses and have facilitated the discovery of new materials that exhibit hyperbolic behavior. Additionally, improvements in nano-optical techniques now enable the probing of high wavenumber regions of the frequency-wavenumber plane, uncovering quantum properties of polaritons. 

In our current formalism, electron-electron interactions are taken into account through a phenomenological approach (see, for example, Ref.~\onlinecite{Cardoso:2025}). As an exciting avenue for future research, we propose integrating many-body phases, such as superconductivity in twisted bilayer graphene, into the hydrodynamic formalism. Exploring other highly correlated phases within this framework presents a compelling challenge.

\section*{Acknowledgments}

N.~M.~R.~P. acknowledges support by the Portuguese Foundation for Science and Technology (FCT) in the framework of the projects PTDC/FIS-MAC/2045/2021 and EXPL/FISMAC/0953/2021, and the Strategic Funding UIDB/04650/2020. L.~J. and J.~D.~C. acknowledge support from Independent Research Fund Denmark (grant no. 0165-00051B). N.~M.~R.~P. and N.~A.~M. also acknowledge the Independent Research Fund Denmark (grant no. 2032-00045B).
D.~R.~C. and A.~J.~C. acknowledge the Brazilian Council for Research (CNPq) grants No. $423423/2021-5$, $408144/2022-0$. D.~R.~C. gratefully acknowledges the support from CNPq grants $313211/2021-3$, $437067/2018-1$, and the Research Foundation--Flanders (FWO). A.~J.~C. was supported by CNPq Grant No. 315408/2021-9, FAPESP under Grant No.~2022/08086-0, and the Brazilian National Council for the Improvement of Higher Education (CAPES) under a CAPES-PrInt scholarship. The Center for Polariton-driven Light--Matter Interactions (POLIMA) is sponsored by the Danish National Research Foundation (Project No.~DNRF165).

\appendix

\section{Fresnel coefficients for a anisotropic 2D sheet}\label{app:loss}

We consider light impinging on a 2D sheet described by the surface conductivity tensor $\sigma$. We define the TM and TE direction versors $\mathbf{e}_p$ and $\mathbf{e}_s$, so that the incident electric field is given by
\begin{equation}
    \mathbf{E}_i=E_{i,p} \mathbf{e}_p e^{i\mathbf{k}\cdot \mathbf{r}} +E_{i,s}\mathbf{e}_s e^{i\mathbf{k}\cdot\mathbf{r}},
\end{equation}
the reflected field component is
\begin{equation}
    \mathbf{E}_r=E_{r,p} \mathbf{e}_p^\prime e^{i\mathbf{k}^ \prime\cdot \mathbf{r}} +E_{r,s}\mathbf{e}_s e^{i\mathbf{k}^ \prime\cdot\mathbf{r}},
\end{equation}
while the transmitted field component one is
\begin{equation}
    \mathbf{E}_t=E_{t,p} \mathbf{e}_p e^{i\mathbf{k}\cdot \mathbf{r}} +E_{t,s}\mathbf{e}_s e^{i\mathbf{k}\cdot\mathbf{r}},
\end{equation}
with $\mathbf{k}=(k_\parallel\cos\phi,k_\parallel\sin\phi,k_z)$ and $\mathbf{k}^\prime = (k_\parallel\cos\phi,k_\parallel\sin\phi,-k_z)$. The unit vectors are given by
\begin{subequations}
\begin{eqnarray}
\mathbf{e}_p &=&\frac{k_z}{k_0}\cos\phi \,\mathbf{e}_x+\frac{k_z}{k_0} \sin\phi\,\mathbf{e}_y-\frac{k_\parallel}{k_0}\,\mathbf{e}_z,\\
\mathbf{e}_s &=&\sin\phi\, \mathbf{e}_x-\cos\phi\, \mathbf{e}_y,\\
\mathbf{e}_p^\prime &=&-\frac{k_z}{k_0}\cos\phi\,\mathbf{e}_x-\frac{k_z}{k_0} \sin\phi\,\mathbf{e}_y-\frac{k_\parallel}{k_0}\mathbf{e}_z,\quad
\end{eqnarray}
\end{subequations}
where we note that $\mathbf{e_p}\cdot\mathbf{e}_s=\mathbf{e_p}^\prime\cdot\mathbf{e}_s=0$.

The corresponding magnetic field for each component is written as
\begin{subequations}
\begin{eqnarray}
    \mathbf{B}_i=-\frac{1}{c} E_{i,p} \mathbf{e}_s e^{i\mathbf{k}\cdot \mathbf{r}} +\frac{1}{c}E_{i,s}\mathbf{e}_p e^{i\mathbf{k}\cdot\mathbf{r}},\\
    \mathbf{B}_r =-\frac{1}{c} E_{r,p} \mathbf{e}_s e^{i\mathbf{k}^ \prime\cdot \mathbf{r}} +\frac{1}{c}E_{r,s}\mathbf{e}_p^\prime e^{i\mathbf{k}^ \prime\cdot\mathbf{r}},\\
    \mathbf{B}_t =-\frac{1}{c} E_{t,p} \mathbf{e}_s e^{i\mathbf{k}\cdot \mathbf{r}} +\frac{1}{c}E_{t,s}\mathbf{e}_p e^{i\mathbf{k}\cdot\mathbf{r}}.
\end{eqnarray}
\label{eq:Bfields}
\end{subequations}
From the continuity of the electric field at the 2D material interface, we have:
\begin{subequations}
\begin{eqnarray}
E_{i,p}-E_{r,p}=E_{t,p},\\
E_{i,s}+E_{r,s}=E_{t,s}.
\end{eqnarray}
\label{eq:boundary}
\end{subequations}

From {\`A}mpere's law at the interface,
\begin{equation}
(\mathbf{H}_t-\mathbf{H}_i-\mathbf{H}_r)\times\mathbf{e}_z
=\sigma \mathbf{E}_{t,\parallel},
\end{equation}
we now decompose in $p$ and $s$ modes, where we use that
\begin{subequations}
    \begin{flalign}
\left(\mathbf{e}_p\times\mathbf{u}_z\right)\cdot\mathbf{e}_s
=\frac{k_z}{k_0}, \\
\left(\mathbf{e}_s\times\mathbf{u}_z\right)\cdot\mathbf{e}_p
=-\frac{k_z}{k_0}.
\end{flalign}
\end{subequations}
This implies that
\begin{subequations}
    \begin{multline}
\left[(\mathbf{H}_t-\mathbf{H}_i-\mathbf{H}_r)\times\mathbf{e}_z\right]\cdot \mathbf{e}_p
\\=-\frac{k_z}{k_0}\left( H_{t,s}-H_{r,s}-H_{i,s} \right),
\end{multline}
and
\begin{multline}
\left[(\mathbf{H}_t-\mathbf{H}_i-\mathbf{H}_r)\times\mathbf{e}_z\right]\cdot \mathbf{e}_s
\\=\frac{k_z}{k_0}\left( H_{t,p}+H_{r,p}-H_{i,p} \right).
\end{multline} \label{eq:Hp_Hs}
\end{subequations}

Next, we define the conductivity in a rotated frame
\begin{equation}
\tilde{\sigma}(\phi)=R^T_\phi \begin{pmatrix} \sigma_{xx} & \sigma_{xy} \\ \sigma_{yx} & \sigma_{yy}\end{pmatrix} R_\phi
\end{equation}
with the rotation matrix
\begin{equation}
R_\phi=\begin{pmatrix}
    \cos\phi & -\sin\phi \\ \sin\phi & \cos\phi
\end{pmatrix}.
\end{equation}
With this, we have
\begin{eqnarray}
    \mathbf{J}\cdot\mathbf{e}_p= \frac{k_z^2}{k_0^2} \tilde{\sigma}_{xx}(\phi)E_p
    +\frac{k_z}{k_0}\tilde{\sigma}_{xy}(\phi)E_s,\\
    \mathbf{J}\cdot\mathbf{e}_s= \frac{k_z}{k_0} \tilde{\sigma}_{yx}(\phi)E_p
    + \tilde{\sigma}_{yy}(\phi)E_s,
\end{eqnarray}
which leads to the following set of equations:
\begin{multline}
-\frac{k_z}{k_0}\left( H_{t,s}-H_{r,s}-H_{i,s} \right)\\=
\frac{k_z^2}{k_0^2} \tilde{\sigma}_{xx}^\phi E_{t,p}
    +\frac{k_z}{k_0}\tilde{\sigma}_{xy}^\phi E_{t,s},
\end{multline}
\begin{multline}
\frac{k_z}{k_0}\left( H_{t,p}+H_{r,p}-H_{i,p} \right)\\=\frac{k_z}{k_0} \tilde{\sigma}_{yx}^\phi E_{t,p}
    + \tilde{\sigma}_{yy}^\phi E_{t,s}.
\end{multline}

From Eqs.~\eqref{eq:Bfields} and~\eqref{eq:Hp_Hs}, we have
\begin{eqnarray}
B_{i,p}=\frac{1}{c} E_{i,s}, \\
B_{i,s}=-\frac{1}{c}E_{i,p},
\end{eqnarray}
and thus also
\begin{multline}
 E_{t,p}-E_{r,p}-E_{i,p} \\=\mu_0c\left(
\frac{k_z}{k_0} \tilde{\sigma}_{xx}E_{t,p}
    +\tilde{\sigma}_{xy}E_{t,s}\right),
\end{multline}
\begin{multline}
 E_{t,s}+E_{r,s}-E_{i,s} \\= \mu_0c\left(\tilde{ \sigma}_{yx}E_{t,p}
    +\frac{k_0}{k_z} \tilde{\sigma}_{yy}E_{t,s}\right).
\end{multline}
Using Eq.~\eqref{eq:boundary}, we in turn obtain
\begin{multline}
2\left( E_{t,p}-E_{i,p} \right)\\=
\frac{k_z }{k_0 } \mu_0c\tilde{\sigma}_{xx} E_{t,p}
    + \mu_0c\tilde{\sigma}_{xy} E_{t,s},
\end{multline}
\begin{multline}
2\left( E_{t,s}-E_{i,s} \right)\\= \mu_0c\tilde{\sigma}_{yx} E_{t,p}
    + \frac{ k_0}{k_z}\mu_0c\tilde{\sigma}_{yy} E_{t,s},
\end{multline}
which can be conveniently rearranged to 
\begin{subequations}
\begin{multline}
\left( \frac{2k_z}{k_0} -\frac{k_z^2}{k_0^2} \mu_0c\tilde{\sigma}_{xx}\right)E_{t,p}
-\frac{k_z}{k_0}\mu_0c\tilde{\sigma}_{xy}E_{t,s}\\=
 \frac{2k_z}{k_0}E_{i,p},
\end{multline}
\begin{multline}
-\frac{k_z}{k_0} \mu_0c\tilde{\sigma}_{yx} E_{t,p}+
\left(\frac{2k_z}{k_0} -\mu_0c\tilde{\sigma}_{yy} \right)E_{t,s}\\=
    \frac{2k_z}{k_0}E_{i,s}.
\end{multline} \label{eq:linear_scatt}
\end{subequations}

Next, we define
\begin{multline}
    D^\prime=\left( \frac{2k_z}{k_0} -\frac{k_z^2}{k_0^2} \mu_0c\tilde{\sigma}_{xx}\right)
\left(\frac{2k_z}{k_0} -\mu_0c\tilde{\sigma}_{yy} \right)
\\-\left(\frac{k_z}{k_0}\mu_0c\right)^2\tilde{\sigma}_{xy}\tilde{\sigma}_{yx}.
\end{multline}
The solution of Eqs.~\eqref{eq:linear_scatt} now becomes
 \begin{multline}\begin{pmatrix}
E_{t,p} \\
E_{t,s}
\end{pmatrix}=
\frac{2k_z^2}{k_0^2D^\prime
}
\\\times
\begin{pmatrix}
 2 -\frac{k_0}{k_z}\mu_0c\tilde{\sigma}_{yy} & \mu_0c\tilde{\sigma}_{xy}\\
 \mu_0c\tilde{\sigma}_{yx} & \frac{2 }{ } -\frac{k_z }{k_0 } \mu_0c\tilde{\sigma}_{xx}
\end{pmatrix}
\begin{pmatrix} 
 E_{i,p}
\\
 E_{i,s}
\end{pmatrix}, \label{eq:i_to_t}
\end{multline}
which in turn implies the following transmission coefficients
\begin{subequations}
\begin{eqnarray}
t_{p,p}&=&\frac{2k_z}{k_0} \frac{\frac{2k_z}{k_0}-\mu_0c\tilde{\sigma}_{yy}}{D^\prime
}, \\
t_{p,s}&=&\frac{2k_z}{k_0} \frac{\frac{k_z}{k_0}\mu_0c\tilde{\sigma}_{xy}}{D^\prime},
\\
t_{s,p}&=&
\frac{2k_z}{k_0} \frac{\frac{k_z}{k_0}\mu_0c\tilde{\sigma}_{xy}}{D^\prime},
\\
t_{s,s}&=&
\frac{2k_z}{k_0} \frac{\frac{2k_z}{k_0}-\frac{k_z^2}{k_0^2}\mu_0c\tilde{\sigma}_{xx}}{D^\prime}.
\end{eqnarray}
\end{subequations}
Likewise, for the reflection part 
\begin{equation}
\begin{pmatrix}
E_{r,p} \\ E_{r,s}
\end{pmatrix}=\begin{pmatrix}
r_{p,p} & r_{s,p} \\ 
r_{p,s} & r_{s,s}
\end{pmatrix} \begin{pmatrix}
    E_{i,p} \\ E_{i,s}
\end{pmatrix},
\end{equation}
the use of Eqs.~\eqref{eq:boundary} formally gives
\begin{eqnarray}
r_{p,p}&=&1-t_{p,p}, \\
r_{p,s}&=&t_{p,s},\\
r_{s,p}&=&-t_{s,p},\\
r_{s,s}&=&-1+t_{s,s}.
\end{eqnarray}
Combining these results, we now explicitly have
\begin{multline}
r_{p,p}=\frac{1}{D^\prime}\frac{k_z^2}{k_0^2}
 \Bigg[
    \mu_0c\tilde{\sigma}_{xx} 
\left(\mu_0c\tilde{\sigma}_{yy} -\frac{2k_z}{k_0} \right)
\\-\left(\mu_0c\right)^2\tilde{\sigma}_{xy}\tilde{\sigma}_{yx}
\Bigg].
\end{multline}
In a similar way, we obtain that
\begin{multline}
r_{s,s}=\frac{1}{D^\prime}\left[
 \left(\frac{2k_z}{k_0} -\frac{k_z^2}{k_0^2} \mu_0c\tilde{\sigma}_{xx} \right)
 \mu_0c\tilde{\sigma}_{yy} \right. \\ \left. 
 -\left(\frac{k_z}{k_0}\mu_0c\right)^2\tilde{\sigma}_{xy}\tilde{\sigma}_{yx}
\right].
\end{multline}

\section{Hydrodynamic optical conductivity: the case of losses} \label{app:losses}

In the presence of losses, the Euler equation~\eqref{eq:Euler} can be modified by replacing the time derivative as $\partial_t\rightarrow \partial_t+\gamma$, where $\gamma$ represents a phenomenological damping rate~\cite{Cardoso:2025}:
 \begin{multline}
\left(\partial_t+i\gamma\right) u_k+ 
\sum_j u_j\partial_j u_k \\
+\frac{e}{ m_k}\sum_j u_j \left( \partial_kA_j- \partial_jA_k \right)
\\ = -\frac{\partial_k}{m_k}\left[
V^B+U_\mathrm{ext}-e(\Phi+\partial_tA_k)
\right]. \label{eq:Euler_damped}
\end{multline}
Proceeding along the procedure in Subsec.~\ref{subsec:nonlocal_cond} we have
\begin{multline}
M\left(\partial_t +\gamma\right)\mathbf{u} +e\mathbf{u}_1\times\mathbf{B}
= \frac{\hbar^2}{4}\sum_j \frac{1}{m_j } \frac{\partial_j^2\mathbf{\nabla} n }{n_0}\\-e\mathbf{E}-\frac{\hbar^2 \pi}{ \sqrt{m_xm_y}} \mathbf{\nabla} n , \label{eq:lin-loss}
\end{multline}
and we eventually arrive at the nonlocal optical conductivity tensor
\begin{multline}
\sigma(\mathbf{q},\omega)=\frac{\sigma_0(\omega) }{D_\gamma(\omega)}\frac{m_0}{m_x m_y} \\ \times \begin{pmatrix}
 m_y(\omega+i\gamma)\omega - q_y^2 f_\mathbf{q}& -ieB_z\omega+ q_xq_y f_\mathbf{q} \\
 i e B_z\omega+ q_xq_y f_\mathbf{q} & m_x(\omega+i\gamma)\omega -q_x^2f_\mathbf{q}
\end{pmatrix}, \label{eq:cond_fin-loss}
\end{multline}
where we have defined
\begin{equation}
    D_\gamma(\omega)=\left(\omega+i\gamma\right)^2\omega-\omega^2_\mathbf{q}\omega -i\gamma f_\mathbf{q}\left(\frac{q_x^2}{m_x}+\frac{q_y^2}{m_y}\right).
\end{equation}

\section{dyadic Green's function matrices} \label{app:dya}

For completeness, we present the matrices $M_{ij}$ of Eq.~\eqref{eq:dyadic}:
\begin{equation}
 M_{ss}=\frac{1}{k_z k_\parallel^2}
 \begin{pmatrix}
     k_y^2 & -k_xk_y & 0 \\
     -k_xk_y & k_x^2 & 0 \\
     0 & 0 & 0
 \end{pmatrix},
\end{equation}
\begin{equation}
 M_{sp}=\frac{1}{k_0 k_\parallel^2}
 \begin{pmatrix}
     -k_x k_y & -k_y^2 & -k_yk_\parallel^2/k_z \\
     k_x^2 & k_xk_y & k_xk_\parallel^2/k_z \\
     0 & 0 & 0
 \end{pmatrix},
\end{equation}
\begin{equation}
 M_{ps}=\frac{1}{k_0 k_\parallel^2}
 \begin{pmatrix}
     k_xk_y & -k_x^2 & 0 \\
     k_y^2 & -k_xk_y & 0 \\
     -k_yk_\parallel^2/k_z & k_xk_\parallel^2/k_z & 0
 \end{pmatrix},
\end{equation}
\begin{equation}
 M_{pp}=\frac{k_z}{k_0^2 k_\parallel^2}
 \begin{pmatrix}
     -k_x^2 & -k_xk_y & -k_xk_\parallel^2/k_z \\
     -k_xk_y & k_y^2 & -k_yk_\parallel^2/k_z \\
     k_xk_\parallel^2/k_z & k_yk_\parallel^2/k_z & k_\parallel^4/k_z^2
 \end{pmatrix}.
\end{equation}
For further details, we refer to the derivation in Ref.~\onlinecite{Gomez-Diaz2015}.

\section{Surface Plasmon-Polariton dispersion relation} \label{app:spp_disp}

In this section, we solve Eq.~\eqref{eq:new_pole} in the limit where Bohm's potential is neglected:
\begin{equation}
2-\frac{k_z}{k_0}\mu_0 c \tilde{\sigma}_{xx}=0. \label{eq:TM_disp}
\end{equation}
The conductivity in the rotated frame is now 
\begin{multline}
\mu_0 c\tilde{\sigma}_{xx} = \mu_x \mu_y u_\phi \frac{A}{1-C_ps^2u_\phi}
\\+k_0d(1-\epsilon_x\cos^2\phi-\epsilon_y^2\sin^2\phi), \label{eq:cond_xx}
\end{multline}
where we have defined the constants
\begin{subequations}
\begin{eqnarray}
A=\frac{4\pi \alpha \tilde{\lambda}_c n_0}{k_0\mu_x\mu_y},\\
C_p=
\frac{\pi n_0 \tilde{\lambda}_c^2}{\sqrt{\mu_x\mu_y}},
\end{eqnarray}
and the function 
\begin{equation}
    u_\phi = \frac{\cos^2\phi}{\mu_x}+\frac{\sin^2\phi}{\mu_y}.
\end{equation}
\end{subequations}
By substituting Eq.~\eqref{eq:cond_xx} into Eq.~\eqref{eq:TM_disp} and carrying out some algebraic manipulations, the resulting expression can be recast as a quadratic equation in $\cos(2\phi)$:
\begin{widetext}
\begin{multline}
     \Big\{C_p s^2 \left(\tfrac{1}{\mu_y}-\tfrac{1}{\mu_x}\right)k_0d\tfrac{\epsilon_y-\epsilon_x}{4}\Big\}\cos^2(2\phi)\\+
    \Big\{\tfrac{\mu_y-\mu_x}{2}A 
    -\tfrac{1}{2}C_ps^2\left(\tfrac{1}{\mu_x}-\tfrac{1}{\mu_y} \right)\left[k_0d\left(1-\tfrac{\epsilon_x+\epsilon_y}{2} \right)-\tfrac{2}{s} \right]
    +\left[1-\tfrac{1}{2}C_ps^2\left(\tfrac{1}{\mu_x}+\tfrac{1}{\mu_y}\right)\right]k_0d \tfrac{\epsilon_y-\epsilon_x}{2} \Big\}\cos(2\phi)\\
    +\Big\{\tfrac{\mu_y+\mu_x}{2}A +\left[1-\tfrac{1}{2}C_ps^2\left(\tfrac{1}{\mu_x}+\tfrac{1}{\mu_y} \right)\right]
    \left[k_0d\left(1-\tfrac{\epsilon_x+\epsilon_y}{2} \right)-\tfrac{2}{s} \right]\Big\} =0.
\end{multline}
\end{widetext}
Being a second-order polynomial, it is straightforward to obtain the two possible roots. The possible angles are limited by $|\cos2\phi|\le 1$.

\section{Numerical calculation of the Purcell Factor} \label{app:Purcell}

We start rewriting Eq.~\eqref{eq:purcell} in polar coordinates:
\begin{equation}
\frac{\Gamma}{\Gamma_0}= 1 +\frac{3}{4\pi k_0^3} \Im\left[\iint d{k}_\parallel d\phi\frac{k_\parallel^3e^{2i\sqrt{k_0^2-k_\parallel^2}z_0} }{\sqrt{k_0^2-k_\parallel^2}} r_{pp}\right]. \label{eq:purcell_polar}
\end{equation}
Next, we split the integral into the regions $k_\parallel<k_0$ and $k_\parallel>k_0$. For the first region, we have
\begin{multline}
I=\Im\left[\int_0^{k_0} d{k}_\parallel \int_0^{2\pi}d\phi\frac{k_\parallel^3e^{2i\sqrt{k_0^2-k_\parallel^2}z_0} }{\sqrt{k_0^2-k_\parallel^2}} r_{pp}\right]\\
\le 
\int_0^{k_0} d{k}_\parallel \int_0^{2\pi}d\phi \left|\frac{k_\parallel^3e^{2i\sqrt{k_0^2-k_\parallel^2}z_0} }{\sqrt{k_0^2-k_\parallel^2}} r_{pp}\right|, 
\end{multline}
and using that $|r_{pp}|<1$, we have
\begin{equation}
    I\le \int_0^{k_0}dk_\parallel \int_0^{2\pi} d\phi \frac{k_\parallel^3}{\sqrt{k_0^2-k_\parallel^2}}
    =\frac{4\pi k_0^3}{3}.
\end{equation}
Therefore, the contribution to the Purcell factor from the region $k_\parallel<k_0$ cannot exceed unity. The contribution from the region $k_\parallel>k_0$ is several orders of magnitude larger; therefore, the term $I$ can be conveniently neglected in the numerical calculation of the Purcell factor,
\begin{equation}
\frac{\Gamma}{\Gamma_0}\approx 1 +\frac{3}{4\pi } \int_0^\infty ds (s^2+1)e^{-2sk_0z_0} \int_0^{2\pi}d\phi\, \Im\left[r_{pp}\right], \label{eq:purcell_polar2}
\end{equation}
where the dimensionless integration variable is defined as $s=\sqrt{k_\parallel^2-k_0^2}/k_0$. This expression was used to obtain the Purcell factors of Fig.~\ref{fig:dipoles}\pnl{c}.

\bibliographystyle{apsrev4-1}
\bibliography{references.bib}

\end{document}